\documentclass [12pt,eqsecnum,amsfonts,aps,amsymb]{revtex4}

\input epsf

\newcommand{\beq}{\begin{equation}}
\newcommand{\eeq}{\end{equation}}
\newcommand{\beqs}{\begin{eqnarray}}
\newcommand{\eeqs}{\end{eqnarray}}

\begin{document}

\baselineskip 6.0mm

\title{Exact Partition Functions for the $q$-State Potts Model with a 
Generalized Magnetic Field on Lattice Strip Graphs} 

\author{Shu-Chiuan Chang$^{a,b}$ \footnote{(a): Permanent
    address; (b) Address on sabbatical} and Robert Shrock$^b$}

\affiliation{(a) \ Department of Physics, National Cheng Kung University,
Tainan 70101, Taiwan}

\affiliation{(b) \ C. N. Yang Institute for Theoretical Physics and
Department of Physics and Astronomy \\
Stony Brook University, Stony Brook, NY 11794, USA }

\begin{abstract}

We calculate the partition function of the $q$-state Potts model on
arbitrary-length cyclic ladder graphs of the square and triangular lattices,
with a generalized external magnetic field that favors or disfavors a subset of
spin values $\{1,...,s\}$ with $s \le q$. For the case of antiferromagnet
spin-spin coupling, these provide exactly solved models that exhibit an onset
of frustration and competing interactions in the context of a novel type of
tensor-product $S_s \otimes S_{q-s}$ global symmetry, where $S_s$ is the
permutation group on $s$ objects.

\end{abstract}

\pacs{}

\maketitle


\pagestyle{plain}
\pagenumbering{arabic}


\section{Introduction}
\label{intro}

We consider the $q$-state Potts model in a generalized external magnetic field
that favors or disfavors a certain subset of spin values in the interval $I_s =
\{1,...,s\}$ with $s \le q$.  We present exact solutions for the partition
function of this model on ladder strips of the square and triangular lattices
with arbitrary length and periodic (i.e., cyclic) boundary conditions in the
longitudinal ($x$) direction and free boundary conditions in the transverse
($y$) direction. This work extends our previous calculations of the partition
function for the cyclic ladder graph in the zero-field case in
Refs. \cite{a,ta} and for the case of nonzero field and $s=1$ in Ref.
\cite{zth}.

In general, we denote a graph $G=(V,E)$ by its vertex set $V$ and its edge ( =
bond) set $E$.  The numbers of vertices, edges, and connected components of $G$
are denoted, respectively, by $n(G) \equiv n$, $e(G)$, and $k(G)$. We note
that a triangular-lattice ladder graph of a given length may be obtained from
the square-lattice ladder graph by adding a diagonal edge in each square,
connecting, say, the lower left vertex to the upper right vertex in the
square. In thermal equilibrium at temperature $T$, the partition function for
the $q$-state Potts model on the graph $G$ in this generalized external
magnetic field is given by
\beq
Z = \sum_{ \{ \sigma_i \} } e^{-\beta {\cal H}} \ , 
\label{z}
\eeq
with the Hamiltonian
\beq
{\cal H} = -J \sum_{\langle i j \rangle} \delta_{\sigma_i, \sigma_j}
- \sum_{p=1}^q H_p \sum_i \delta_{\sigma_i,p} \ ,
\label{ham}
\eeq
where $i$ and $j$ label vertices of $G$, $\sigma_i$ are
spin variables with values in the set 
$I_q = \{1,...,q\}$, $\beta = (k_BT)^{-1}$, $\langle
i j \rangle$ denote pairs of adjacent vertices, and $J$ is the spin-spin
interaction constant.  We denote the subset of the $q$ spin values
orthogonal to $I_s$ as $I_s^\perp = \{s+1,...,q\}$. 
The generalized magnetic field is defined as 
\beq
H_p = \cases{ H & for $p \in I_s$ \cr
              0 & for $p \in I_s^\perp$ } \ .
\label{hp}
\eeq
Hence, for $H > 0$ (respectively $H < 0$), this model favors (resp. disfavors)
spin values in the interval $I_s$.  As is evident in Eqs. (\ref{z}) and
(\ref{ham}), the spin-spin interaction constant $J$ and the external magnetic
field $H$ always appear in $Z$ multiplied by $\beta$, and, furthermore, in
exponentiated form, so it is natural to define the following quantities:
\beq
K = \beta J \ , \quad h = \beta H \ , \quad y = e^K \ , \quad v = y-1 \ ,
\quad w=e^h \ .
\label{kdef}
\eeq
The physical ranges of $v$ are $v \ge 0$ for the Potts ferromagnet, and
$-1 \le v \le 0$ for the Potts antiferromagnet.

There are several motivations for this work. One motivation is the insight that
it gives into how a new type of global symmetry is manifested in the structure
of the partition function.  If and only if $H=0$, then the zero-field Potts
model Hamiltonian ${\cal H}$ and partition function $Z$ are invariant under the
global symmetry group in which $\sigma_i \to g \sigma_i \ \forall \ i \in V$,
with $g \in S_q$, where $S_q$ is the symmetric (= permutation) group on $q$
objects.  We have used this symmetry property to choose, without loss of
generality, the set of spins $I_s$ to be contiguous.  Turning on a conventional
external magnetic field that favors (or disfavors) one value of the $\sigma_i$
has the effect of simply reducing the original $S_q$ permutation symmetry group
to $S_{q-1}$. (In the $q=2$ Ising case, this removes any global symmetry,
since $S_1$ is trivial.) In contrast, with the generalized external magnetic
field here, setting $H \ne 0$, reduces the original $S_q$ global permutation
symmetry to the novel type of tensor-product global symmetry group
\beq
{\cal G}_{sym} = S_s \otimes S_{q-s} \ .
\label{symmetrygroup}
\eeq
This tensor-product symmetry group in the presence of $H \ne 0$ has interesting
consequences for the structure of the partition function, as will be evident
from several identities to be discussed below.  

A second motivation for the present work is the connection with certain
polynomials that appear in mathematical graph theory.  In the case of zero
external magnetic field, the partition function of the $q$-state Potts model is
equivalent to a function of central interest in modern mathematical graph
theory, namely the Tutte polynomial. If the spin-spin exchange constant, $J$,
is negative, i.e., antiferromagnetic, the spin-spin interaction favors spin
configurations such that spins on adjacent vertices are different.  In the
limit of this Potts antiferromagnet in which the temperature $T \to 0$ (i.e.,
$v \to -1$), the only spin configurations that make a nonzero contribution to
the partition function are those such that spins on adjacent vertices have
different values.  It follows that the partition function for this
zero-temperature $q$-state Potts antiferromagnet is equal to the chromatic
polynomial $P(G,q)$, which counts the number of ways of assigning $q$ colors to
the vertices of the graph $G$ subject to the condition that the colors on
adjacent vertices are different. Such a color assignment is called a proper
$q$-coloring of (the vertices of) $G$. Our generalization to a nonzero external
magnetic field that favors or disfavors a subset of spin values establishes a
further connection with mathematical graph theory.  This connection is embodied
in the purely graph-theoretic expression that we presented (as Eq. (12.7) in
\cite{ph}) for the partition function.  The zero-temperature limit of this
partition function in the antiferromagnetic case defined what we have called a
weighted-set graph coloring problem. Although we will not pursue applications
here, we have discussed these earlier \cite{ph}.  For example, the weighted-set
graph coloring problem could describe the assignment of $q$ frequencies to
commercial radio broadcasting stations in an area such that adjacent stations
must use different frequencies to avoid interference and (i) for $H > 0$, the
stations prefer frequences in a set $I_s$ since reception is better for these
or (ii) for $H < 0$, the stations prefer to avoid the frequencies in the 
set $I_s$ because of poorer reception.

A third motivation is that our solutions provide exactly solved examples that
exhibit the presence or absence of frustration and competing interactions in a
controlled, parameter-dependent manner. To explain this, we note that for a
graph $G$, the minimum number of colors necessary to carry out a proper
$q$-coloring of (the vertices of) $G$ is the chromatic number of $G$, denoted
$\chi(G)$.  If the spin-spin exchange interaction, $J$, is negative, then the
minimization of the internal energy favors spin configurations in which the
spins on adjacent vertices have different values.  We consider this situation,
$J < 0$, for the remainder of this paragraph.  Now in the presence of a nonzero
external magnetic field $H$ that is positive (respectively, negative), the
minimization of the internal energy favors spin values that lie in the interval
$I_s$ (resp. $I_s^\perp$).  The question of whether or not either (or both) of
these two different situations involves competing interactions and associated
frustration depends on the value of $s$ or ($q-s$) relative to the chromatic
number $\chi(G)$ of $G$.  In the rest of this paragraph, we assume that $s \ne
q$ since if $s=q$, then the partition function can be expressed completely (see
Eq. (\ref{zsq}) below) in terms of the partition function of a zero-field model
multiplied by an overall factor.  We also assume that $s$ does not have the
formal value $s=0$, since then the $H$ field has no effect (see Eq. (\ref{zs0})
below).  Now, if $H > 0$ and $0 < s < \chi(G)$ or if $H < 0$ and $0 \le q-s <
\chi(G)$, then the application of the respective positive or negative $H$
involves competing interactions and causes frustration.  The reason for this
can be explained as follows.  If $H > 0$ and $s < \chi(G)$, then the
restriction preferred by the magnetic field interaction, that the spins should
take values in $I_s$, conflicts with the restriction preferred by the spin-spin
interaction, that adjacent spins should be different, requiring that they take
on values in the set $ \{ 1,..., \chi(G) \}$, which is larger than the interval
$I_s$.  The same statement applies for negative $H$, with the replacement of
$s$ by $q-s$ and thus $I_s$ by $I_s^\perp$. This frustration becomes especially
strong as the temperature $T \to 0$, i.e., $K \to \-\infty$ and $|h| \to
\infty$ and for the infinite strip it leads to a nonanalytic change in the
properties of the ground state as a function of $H/|J|$ if $J < 0$ and $s <
\chi(G)$.  A particularly useful aspect of the present work is that one can
see the origin of this nonanalyticity explicitly from the structure of the
exactly solved free energy.  (For other examples of 
frustration and competing interactions in spin models, see, e.g., 
\cite{salinas}.) 

More generally, an obvious motivation for exact calculations of partition
functions of statistical mechanical models is the insights that they can give
concerning the properties of these models that complement those obtained
from numerical simulations and series expansions.

This paper is organized as follows.  In Sect. \ref{properties} we review basic
properties of $Z(G,q,s,v,w)$ for an arbitrary graph $G$.  In
Sect. \ref{generalstrip} we recall some general structural results for
$Z(G,q,s,v,w)$ where $G$ is an arbitrarily long cyclic strip graph of the
square or triangular lattice that will be relevant for our specific
calculations. In Sects.  \ref{sqladder} and \ref{triladder} we present our
calculations of this partition function for arbitrarily long ladder graphs of
the square and triangular lattices, respectively.  In Sect. \ref{therm} we
discuss some physical consequences of these calculations.
Sect. \ref{conclusions} contains our conclusions, and several relevant formulas
are given in the Appendices \ref{appendix_zlad}-\ref{appendix_tztrid10}. 


\section{Basic Properties}
\label{properties}

In this section we will briefly review some basic properties of the partition
function that are relevant for the present work.  Recall that a spanning
subgraph $G' \subset G$ is a graph consisting of the same set of vertices and a
subset of the edges of $G$, i.e., $G'=(V,E')$ with $E' \subset E$.  It is quite
useful to have a formula that expresses $Z$ in a purely graph-theoretic manner,
instead of as a sum over the spin configurations $\sigma_i$. This has been
achieved by expressing $Z$ as a sum of contributions from the disjoint
components $G_i$ comprising spanning subgraphs $G' \subseteq G$. The simplest
formulation of a $q$-state Potts model in an external magnetic field $H$ takes
this field to favor or disfavor one spin value out of the $q$ possible values.
In our notation, this is the case $s=1$, i.e., the interval $I_s$ consists of a
single spin value which, without loss of generality, can be taken to be
$\sigma_i=1$.  For this $s=1$ case, a graph-theoretic formula for the partition
function was given in \cite{wu78} (reviewed in \cite{wurev}). In
Refs. \cite{zth} and \cite{hl,ph} we studied this case further and presented a
number of properties of the resultant partition function for various families
of graphs.  As discussed in the introduction, a further generalization is to
take the external magnetic field to favor or disfavor a set of several spin
values in an interval $I_s$ rather than just a single value.  For this $s > 1$
case, we gave a general formula for the partition function for an arbitrary
graph $G$ (as Eq. (12.7)) in Ref. \cite{ph}.  Refs. \cite{phs,phs2} derived a
number of properties of this partition function and reported exact calculations
of it for several families of graphs. Let us denote the number of connected
components of $G$ as $k(G)$ and the connected subgraphs of a spanning subgraph
$G'$ as $G'_i$, $i=1,..,k(G')$. Our formula is \cite{ph}
\beq
Z(G,q,s,v,w) = \sum_{G' \subseteq G} v^{e(G')} \
\prod_{i=1}^{k(G')} \, u_{n(G'_i)}  \ , 
\label{clusterws}
\eeq
where
\beq
u_m = q-s+sw^m = q+s(w^m-1)  \ . 
\label{um}
\eeq
It follows from Eq. (\ref{clusterws}) that $Z$ is a polynomial in the variables
$q$, $s$, $v$, and $w$, and this is indicated in the notation
$Z(G,q,s,v,w)$. Note that Eqs. (\ref{clusterws}) and (\ref{um}) also allow one
to generalize $q$ and $s$ from the positive integers to the real numbers.
(Indeed, studies of zeros of $Z(G,q,s,v,w)$ in $q$ and/or $s$ for fixed $v$ and
$w$ require that one generalize $q$ and $s$ further to the complex numbers
\cite{ph,phs,phs2}.) Without loss of generality, we restrict ourselves here to
connected graphs $G$; however, the spanning subgraphs $G'$ in Eq.
(\ref{clusterws}) may contain (and most do contain) more than one
connected component.

The case $v=-1$ yields a weighted-set chromatic polynomial, denoted
$Ph(G,q,s,w)$, that describes a graph coloring problem in which one assigns $q$
colors to the vertices of a graph such that adjacent vertices have different
colors, with a vertex weighting $w$ that either disfavors or favors a given
subset of $s$ colors contained in the set of $q$ colors \cite{hl}-\cite{ph}.
In our notation,
\beq
Ph(G,q,s,w) \equiv Z(G,q,s,-1,w) \ . 
\label{phz}
\eeq
As a quantity encoding information about a graph, the weighted-set chromatic
polynomial $Ph(G,q,s,w)$ is more powerful than the chromatic polynomial, since
it can distinguish between graphs that are chromatically equivalent, i.e., are
different but have the same chromatic polynomial \cite{ph}-\cite{phs2}.  

The zero-field $q$-state Potts model with partition function $Z(G,q,v)$ is
equivalent to a function of central importance in mathematical graph theory,
the Tutte polynomial \cite{tutte47,tutte67} (early reviews include 
\cite{biggsbook,bollobas}) 
\beq
Z(G,q,v) = (x-1)^{k(G)}(y-1)^{n(G)} \, \sum_{G' \subseteq G} \, 
(x-1)^{k(G')-k(G)}(y-1)^{c(G')} \ ,
\label{zt}
\eeq
where
\beq
x = 1 + \frac{q}{v}
\label{x}
\eeq
and $y$ was given in Eq. (\ref{kdef}), so that $q=(x-1)(y-1)$. In
Eq. (\ref{zt}), $c(G')$ denotes the number of linearly independent circuits
(cycles) in $G$, satisfying $c(G')=e(G')+k(G')-n(G')$. Just as $Ph(G,q,s,w)$
generalizes the chromatic polynomial, so also $Z(G,q,s,v,w)$ generalizes the
Tutte polynomial.  For example, $Z(G,q,s,v,w)$ can distinguish between
different graphs with the same Tutte polynomnial \cite{phs,phs2}.

We next review some basic identities and symmetry properties of $Z(G,q,s,v,w)$.
From the relation $w^m-1=(w-1)\sum_{j=0}^{m-1}w^j$ with $m=n(G_i')$ in
conjunction with Eqs. (\ref{clusterws}) and (\ref{um}), it follows that the
variable $s$ enters in $Z(G,q,s,v,w)$, and $Ph(G,q,s,w)$ only in the
combination $s(w-1)$. Furthermore, $Z(G,q,s,v,w)$ satisfies the following
identities. First, if $h=0$, i.e., $w=1$, then the partition function reduces
to the zero-field expression result, i.e., 
\beq
Z(G,q,s,v,1) = Z(G,q,v) \ .
\label{zw1}
\eeq
An equivalent result is obtained if one formally sets $s=0$, i.e., the set of
spin values that interact with the external field is the null set: 
\beq
Z(G,q,0,v,w) = Z(G,q,v) \ .
\label{zs0}
\eeq
The equivalence of Eqs. (\ref{zw1}) and (\ref{zs0}) is in accord with the
above-mentioned property that $Z(G,q,s,v,w)$ depends on the variables $s$ and
$w$ only via the combination $s(w-1)$, so $w=1$ is equivalent to $s=0$.  Since
the right-hand sides of Eqs. (\ref{zw1}) and (\ref{zs0}) do not depend on $s$
or $w$, they also imply the partial differential equations
\beq
\frac{\partial Ph(G,q,s,1)}{\partial s} = 0
\label{dzdsw1}
\eeq
and
\beq
\frac{\partial Ph(G,q,0,w)}{\partial w} = 0 \ .
\label{dzdww1}
\eeq
The condition $w = 0$ means that spin configurations with $\sigma_i \in I_s$
for any $i$ make no contribution to the partition function, leading to the
identity 
\beq
Z(G,q,s,v,0) = Z(G,q-s,v) \ . 
\label{zw0} 
\eeq
If $s=q$, then all of the spin values receive the same weighting due to the
generalized external magnetic field, so that 
\beq
Z(G,q,q,v,w) = w^n \, Z(G,q,v) \ .
\label{zsq}
\eeq
Finally, from the global tensor-product symmetry (\ref{symmetrygroup}) of the
theory in the presence of the generalized magnetic field, there is an identity
involving the interchange of the intervals $I_s$ and $I_{q-s}$ combined with a
reversal in the sign of $h$, namely
\beq
Z(G,q,s,v,w) = w^n \, Z(G,q,q-s,v,w^{-1}) \ . 
\label{zsym}
\eeq
Setting $v=-1$ in these identities (\ref{zw1})-(\ref{zsym}) yields the
corresponding relations for $Ph(G,q,s,w)$; for example, $Ph(G,q,s,w) = w^n \,
Ph(G,q,q-s,w^{-1})$, etc.


\section{Partition Function for Cyclic Strip Graphs}
\label{generalstrip}

In this section we recall some general structural results from \cite{phs} that
apply to $Z(G,q,s,v,w)$ for strip graphs of the square or triangular lattice,
denoted generically as $\Lambda$, of arbitarily great length $L_x=m$ vertices
and transverse width $L_y$ vertices with free transverse boundary conditions
and periodic (cyclic) longitudinal boundary conditions. We denote a strip graph
of this type as $G = \Lambda,L_y \times m,cyc.$.  The partition function
$Z(G,q,s,v,w)$ has the general form
\beq
Z(\Lambda, L_y \times m,cyc.,q,s,v,w)
 = \sum_{d=0}^{L_y} \tilde c^{(d)}(q) \sum_{j=1}^{n_{Zh}(L_y,d,s)}
[\lambda_{Z,\Lambda,L_y,d,j}(q,s,v,w)]^m \ ,
\label{zsumcyc}
\eeq
where
\beq
\tilde c^{(d)}(q) \equiv c^{(d)}(\tilde q) =
\sum_{j=0}^d (-1)^j {2d-j \choose j}\tilde q^{\, d-j}
\label{cdtilde}
\eeq
and $\tilde q = q-s$. The first few $\tilde c^{(d)}(q)$ are $\tilde
c^{(0)}(q)=1$,
\beq
\tilde c^{(1)}(q) = \tilde q -1 = q-s-1 \ , 
\label{cd1tilde}
\eeq
and
\beq
\tilde c^{(2)}(q) =\tilde q^{\ 2}-3\tilde q+1 = q^2-2qs+s^2+3(s-q)+1 \ .
\label{cd2tilde}
\eeq
As is evident in the exact solution for the circuit graph
in \cite{phs} and in our new solutions, certain $\lambda_{Z,\Lambda,L_y,d,j}$ 
terms occur with multiplicities greater than 1, which are polynomials in $s$.
The numbers $n_{Zh}(L_y,d,s)$
were derived for $s=1$ (as Theorem 2.1 with Table 1) in \cite{zth}, and results
for arbitrary $s \in I_s$ were given in \cite{phs}. We have
\beq
n_{Zh}(L_y,L_y,s)=1 \ , 
\label{nzhLyLy}
\eeq
\beq
n_{Zh}(L_y,L_y-1,s)=(s+1)L_y+(L_y-1) \ , 
\label{nzhLyLyminus1}
\eeq
\beq
n_{Zh}(L_y+1,0,s)=(s+1)n_{Zh}(L_y,0,s)+n_{Zh}(L_y,1,s) \ , 
\label{nzhLyd0}
\eeq
and, for $1 \le d \le L_y+1$,
\beq
n_{Zh}(L_y+1,d,s)=n_{Zh}(L_y,d-1,s)+(s+2)n_{Zh}(L_y,d,s)+n_{Zh}(L_y,d+1,s) \ .
\label{nzhLyd}
\eeq
For the present case $L_y=2$, these results yield 
$n_{Zh}(2,2,s)=1$ as a special case of (\ref{nzhLyLy}), 
\beq
n_{Zh}(2,1,s)=2s+3
\label{nzhLy2d1}
\eeq
and
\beq
n_{Zh}(2,0,s)=s^2+2s+2  \ .
\label{nzhLy2d0}
\eeq
The total number of $\lambda_{Z,sq,L_y,d,j}$ terms is 
\beq
N_{Zh,L_y,s} = \sum_{d=0}^{L_y} n_{Zh}(L_y,d,s) \ . 
\label{nzhLytots}
\eeq
If $s=1$, then \cite{hl}
\beq
N_{Zh,L_y,1} = \sum_{j=0}^{L_y} {L_y \choose j} {2j \choose j} \ .
\eeq
For general $s$ \cite{phs}, 
\beq
N_{Zh,L_y,s} = \sum_{j=0}^{L_y} {L_y \choose j} {2j \choose j}s^{L_y-j} \ .
\eeq

For general $s$ and $L_y=2$, Eq. (\ref{nzhLytots}) yields 
\beq
N_{Zh,2,s} = \sum_{d=0}^{2} n_{Zh}(2,d,s) = s^2+4s+6 \ . 
\label{nzh2tots}
\eeq

A general result is that the unique $\lambda_{Z,sq,L_y,d,j}$ for $d=L_y$ (for
which we thus suppress the index $j$) is unique, i.e., has multiplicity 1, and
is given by 
\beq
\lambda_{Z,sq,L_y,L_y,j} \equiv \lambda_{Z,sq,L_y,L_y} = v^{L_y} \ .
\label{lamladLyLy}
\eeq
The $s=1$ special case of this was given (as Eq. (4.4)) in \cite{zth}.

The symmetry (\ref{zsym}) under the replacement of $s$ by $q-s$ and $w$ by
$w^{-1}$ , namely $Z(G,q,s,v,w)=w^n Z(G,q,q-s,v,w^{-1})$, has quite interesting
consequences for the structure of this partition function on cyclic lattice
strip graphs (of arbitrary length). A basic result is that under the
replacement $s \to q-s$, a coefficient $\tilde c^{(d)}(q) = c^{(d)}(\tilde q)$
changes to $c^{(d)}(s)$. Hence, taking into account that the $L_y \times m$
cyclic strip of the square or triangular lattice has $n=L_xL_y=mL_y$ vertices,
the symmetry (\ref{zsym}) is realized as
\beqs
& & Z(\Lambda, L_y \times m,cyc.,q,s,v,w) = 
w^{mL_y} \, Z(\Lambda, L_y \times m,cyc.,q,q-s,v,w^{-1}) \cr\cr
& = & \sum_{d=0}^{L_y} c^{(d)}(s) \sum_{j=1}^{n_{Zh}(L_y,d,q-s)}
[w^{L_y} \, \lambda_{Z,\Lambda,L_y,d,j}(q,q-s,v,w^{-1})]^m \ .
\label{zsumcyc_sqms}
\eeqs
In order for this identity to hold, there must be an interplay between changes
in four different quantities: (i) the numbers $n_{Zh}(L_y,d,s)$ in
Eq. (\ref{zsumcyc}) which become $n_{Zh}(L_y,d,q-s)$ in
Eq. (\ref{zsumcyc_sqms}); (ii) the related multiplicity factors for various
$\lambda_{Z,\Lambda,L_y,d,j}$, which were polynomials in $s$ in
Eq. (\ref{zsumcyc}) and become polynomials in $q-s$ in (\ref{zsumcyc_sqms});
the (iii) the coefficients $\tilde c^{(d)}(q) = c^{(d)}(q-s)$ in
(\ref{zsumcyc}), which become $c^{(d)}(s)$ in (\ref{zsumcyc_sqms}); and (iv)
the terms $(\lambda_{Z,\Lambda,L_y,d,j})^m$, which also change under the
replacements $s \to q-s$, $w \to w^{-1}$, and multiplication by $w^{mL_y}$.  

It is natural to analyze how this symmetry (\ref{zsym}) is realized in the
simplest case, namely for the circuit graph. Here, $n_{Zh}(1,0,s)=s+1$, and
$n_{Zh}(1,1,s)=1$.  The exact partition function for
this graph, $Z(1D; 1 \times m,cyc.,q,s,v,w) \equiv Z(C_m,q,s,v,w)$ for the
circuit graph was given in \cite{phs} and is 
\beq
Z(C_m,q,s,v,w) = \sum_{j=1}^{s+1} [\lambda_{Z,C,0,j}(q,s,v,w)]^m +
\tilde c^{(1)}(q) \, v^m \ , 
\label{zcn}
\eeq
where the $\lambda_{Z,C,d,j}$ terms are as follows. For $d=1$, there is a
single term
\beq
\lambda_{Z,C,1,1}=v \equiv \lambda_{Z,C,1} \ . 
\label{lamcnd1}
\eeq
For $d=0$, the $\lambda_{Z,C,0,j}$ consist of the two roots of a quadratic,
denoted as $\lambda_{Z,C,0,\pm}$, 

\beq
\lambda_{Z,C,0,\pm} = \frac{1}{2} \Big [ A \pm \sqrt{R} \ \Big ] \ ,
\label{lamcnd0plusminus}
\eeq
where
\beq
A = q+s(w-1)+v(w+1) \ , 
\label{ac}
\eeq
\beq
R = A^2-4v(q+v)w \ , 
\label{rc}
\eeq
and, if $s \ge 2$, also the terms 
\beq
\lambda_{Z,C,0,j} = vw \ \  {\rm for} \ \ 3 \le j \le s+1 \ . 
\label{lamvw}
\eeq
That is, provided that $s \ge 2$, the terms $\lambda_{Z,C,0,j}$ are all the
same for $j=3,...,s+1$ and hence can be considered to be a single $\lambda$
with multiplicity $s-1$. Hence, one can re-express
$Z(C_m,q,s,v,w)$ in a different but equivalent form (given as Eq. (5.9) in 
\cite{phs2}), 
\beq Z(C_m,q,s,v,w) = \sum_{j=1}^2 [\lambda_{Z,C,0,j}]^m + (s-1)(vw)^m +
(q-s-1)v^m \ .
\label{zcnexplicit}
\eeq

For this circuit graph, the symmetry (\ref{zsym}) involves the combined effects
of changes in four quantities: (i) the numbers $n_{Zh}(1,1,s)=1$ and
$n_{Zh}(1,0,s)=s+1$, (ii) the multiplicity factors for the various terms
$\lambda_{Z,\Lambda,L_y,d,j}$, in particular the multiplicity factor $s-1$ for
a subset of the $\lambda_{Z,C,0,j}$ terms; (iii) the coefficient $\tilde
c^{(1)}(q) = q-s-1$; and (iv) the terms $(\lambda_{Z,C,d,j})^m$
themselves. Since $\tilde c^{(0)}(q)=1$ and the quadratic roots
$\lambda_{Z,C,0,\pm}$ each occur with multiplicity 1, it follows that they must
transform into $w^{-1}$ times themselves, i.e.,
\beq
s \to q-s \ \Rightarrow \ \lambda_{Z,C,0,\pm} \to w^{-1}\lambda_{Z,C,0,\pm} 
\ ,  
\label{lamd0plusminus_sqms}
\eeq
so that with the inclusion of the factor $w^m$ in the relevant $L_y=1$ case of
Eq.  (\ref{zsumcyc_sqms}), the $[(\lambda_{Z,C,0,+})^m +
  (\lambda_{Z,C,0,-})^m]$ in the $d=0$ sector is transformed into itself. The
other part of the $d=0$ sector is comprised of the term $(s-1)(vw)^m$ coming
from the $m$'th power of $\lambda_{Z,C,0,j}=vw$ with multiplicity $s-1$.  Under
the replacements $s \to q-s$ and $w \to w^{-1}$, this term transforms into
$(q-s-1)(v/w)^m$, and with inclusion of the $w^m$ prefactor, this becomes
$(q-s-1)v^m$.  Under the same replacements $s \to q-s$ and $w \to w^{-1}$, the
$d=1$ term in $Z(C_m,q,s,v,w)$, $\tilde c^{(1)}(q) \, v^m = (q-s-1)v^m$,
transforms into $(s-1)v^m$, so with inclusion of the $w^m$ prefactor, this
becomes $(s-1)(vw)^m$.  Thus, under this replacement, the second and third
terms in $Z(C,q,s,v,w)$ interchange with each other.  This involves what could
be called a ``transmigration'' of terms, in which terms from a sector with
value of $d$ switch with terms in a sector with a different value of $d$.

It is also instructive to see how this exact solution satisfies the identity
(\ref{zw1}) for the special case of zero field, $h=0$, i.e., $w=1$.  In this
case, the $s-1$ $\lambda_{Z,C,0,j}$ with $3 \le j \le s+1$ become equal to
$\lambda_{Z,C,1}=v$. In this $w=1$ special case the quadratic roots reduce to
$q+v$ and $v$. Thus, $(s-1)+1=s$ terms $v^m$ associated with the coefficient
$\tilde c^{(0)}(q)=1$ become equal to, and can be grouped with, the $v^m$ term
associated with the coefficient $\tilde c^{(1)}(q)=q-s-1$, yielding the term
$v^m$ with coefficient $s + (q-s-1)=q-1=c^{(1)}$.  This shows how the
zero-field form of the partition function emerges in this limit.  In \cite{ph},
in the context of the analogous phenomenon for the weighted-set chromatic
polynomial $Ph(G,q,s,w)$, we remarked on this and denoted it as a
``transmigration'' process, in which $\lambda$s in a subset of the $\lambda$s
associated with the coefficient $\tilde c^{(d)}(q)$ become equal to a $\lambda$
associated with a coefficient $\tilde c^{(d')}(q)$ with a different degree
$d'$, so that, taking account of their respective coefficients and
multiplicities, they can be grouped together. The same phenomenon occurs, in a
more complicated manner, for $Z(sq,2,q,s,v,w)$ and $Z(tri,2,q,s,v,w)$.


\section{Partition Function for Cyclic Square-Lattice Ladder Graphs}
\label{sqladder}

Next, we present our exact results for the partition functions of the $q$-state
Potts model in a generalized magnetic field on cyclic ladder graphs of the
square lattices of length $L_x=m$ vertices in the longitudinal direction and
$L_y=2$ vertices in the transverse direction, $Z(sq,2 \times m,cyc.,q,s,v,w)$.
This graph has $n=2L_x$ vertices.  The partition function has the form of
(\ref{zsumcyc}) with $n_{Zh}(2,2,s)=1$ as a special case of (\ref{nzhLyLy}),
and with the $n_{Zh}(2,d,s)$ given in Eq. (\ref{nzhLy2d1}) for $d=1$ and in
Eq. (\ref{nzhLy2d0}) for $d=0$.  As noted above, our new result generalizes the
zero-field solution given in \cite{a} and the solution for the special case
$s=1$ given in \cite{zth}.

We calculate the $\lambda_{Z,sq,2,d,j}s$ to be the following:

\begin{itemize}

\item 

For $d=2$, as the $L_y=2$ special case of (\ref{lamladLyLy}), 
there is a single term
\beq
\lambda_{Z,sq,2,2,1} \equiv \lambda_{Z,sq,2,2} = v^2 \ . 
\label{sqlamlad_d2}
\eeq

\item 

For the $d=1$ sector, i.e., the $\lambda_{Z,sq,2,d,j}$s with coefficient
$\tilde c^{(1)}(q)$, there are $2s+3$ terms. We find the following results: (i)
terms equal to $v^2w$, with multiplicity $2(s-1)$, which thus may be denoted as
\beq
\lambda_{Z,sq,2,1,j}=v^2w \quad {\rm for} \ 1 \le j \le 2(s-1) \ , 
\label{sqlamlad_d1_eq1}
\eeq
(ii), next, the terms $\lambda_{Z,sq,2,1,j}$ with $j=2s-1$ and $j=2s$, which we
denote simply as $\lambda_{Z,sq,2,1,\pm}$: 
\beq
\lambda_{Z,sq,2,1,\pm} = \frac{v}{2}\Big [ A \pm \sqrt{R} \ \Big ] \ , 
\label{sqlamlad_d1_eq2}
\eeq
satisfying 
\beq
\lambda_{Z,sq,2,1,\pm} = v \lambda_{Z,C,1,0,\pm} \ , 
\label{sqlamrelc}
\eeq
where $\lambda_{Z,C,1,0,\pm}$, $A$, and $R$, were given in 
Eqs. (\ref{lamcnd0plusminus})-(\ref{rc}), and finally, 
(iii) the terms $\lambda_{sq,2,1,j}$ with $2s+1 \le j \le
2s+3$, which are the roots of the cubic equation
\beqs
& & \lambda^3 - v \Big [q+s(w-1)+v(v+4+w) \Big ] \, \lambda^2 \cr\cr\
& + & v^3 \Big [q(w+1)+s(w-1)(v+1)+v\{q+1+v(w+1)+4w\} \Big ] \, \lambda 
\cr\cr
& - & (q+v)(1+v)v^5w = 0 \ . 
\label{eqsqd1cubic}
\eeqs
The $\lambda_{Z,sq,2,1,\pm}$ terms in Eq. (\ref{sqlamlad_d1_eq2}) and the three
$\lambda_{Z,sq,2,1,j}$ terms that are the roots of Eq. (\ref{eqsqd1cubic}) each
have multiplicity 1.

\item 

For $d=0$, we have terms $\lambda_{Z,sq,2,0,j}$ with
$1 \le j \le s^2+2s+2$.  Since the ordering of the index $j$ is a convention,
we do not specify it below.  These terms and their multiplicities are 
\beq
(i): \quad v(q+v)w  \quad {\rm with \ multiplicity} \ 1 \ , 
\label{sqlamlad_d0_one}
\eeq
\beq
(ii): \quad (vw)^2 \quad {\rm with \ multiplicity} \ s^2-3s+1 \ , 
\label{sqlamlad_d0_eq1}
\eeq
two quadratic roots, each with multiplicity $s-1$: (iii)
\beq
\lambda_{Z,sq,2,0,\pm} = \frac{vw}{2} \, \Big [ A \pm \sqrt{R} \ \Big ] \ ,
\label{sqlamlad_d0_eq2}
\eeq
which satisfy
\beq
\lambda_{Z,sq,2,0,\pm} = w \, \lambda_{Z,sq,2,1,\pm} = 
                        wv \, \lambda_{Z,C,1,0,\pm} \ ,
\label{sqlamwlam}
\eeq
(iv) the three roots of the following cubic equation, each appearing with 
multiplicity $s-1$: 
\beqs
& & \quad 
 \lambda^3 - vw\Big [ q+s(w-1)+v(1+4w)+ v^2w \Big ] \, \lambda^2 \cr\cr 
& + & (vw)^3 \Big [ 2q+s(w-1)(v+1)+v(q+4+w)+v^2(w+1) \Big ] \, \lambda \cr\cr
& - & (vw)^5(q+v)(1+v) = 0 \ , 
\label{sqlamlad_d0_eq3}
\eeqs
and (v) the five roots of an algebraic equation of degree 5, each with
multiplicity 1.  This quintic equation is the characteristic polynomial of the
transfer matrix $T_{Z,sq,2,0}$ given in Appendix \ref{appendix_tzsqd0}.

\end{itemize}

In the special case $s=1$, Eqs. (\ref{nzhLy2d1}) and (\ref{nzhLy2d0}) yield 
$n_{Zh,2,1}=5$ and $n_{Zh,2,0}=5$, respectively.  For this special 
case we have checked that the $\lambda_{Z,sq,2,1,j}$ and $\lambda_{Z,sq,2,0,j}$
for $j=1,...,5$ agree, respectively, with the eigenvalues that we calculated of
the transfer matrices $T_{Z,sq,2,1}$ in Eq. (4.18) and $T_{Z,sq,2,0}$ in
Eq. (8.1) of \cite{zth}.

It is useful to describe how these $\lambda_{Z,sq,2,d,j}$ terms reduce to the
known results for $Z(sq,2 \times m,q,v)$ in the zero-field case presented in
\cite{a} (listed, for reference, in Appendix \ref{appendix_zlad}). To
distinguish the $\lambda_{sq,2,d,j}$ terms in the zero-field partition function
$Z(sq,2 \times m,q,v)$ from the $\lambda_{Z,sq,2,d,j}$ terms in the
field-dependent partition function $Z(sq,2\times m,q,s,v,w)$, we suppress the
subscript $Z$ in the former.  Among the $\lambda_{Z,sq,2,d,j}$ with $d=1$,
first, the $\lambda_{Z,sq,2,1,j}$ with $1 \le j \le 2(s-1)$ obviously reduce to
$v^2=\lambda_{sq,2,2}$. The quadratic roots $\lambda_{Z,sq,1,j}$ with $j=2s-1$
and $2s$ reduce, respectively as
\beq
\lambda_{Z,sq,2,1,+} = v(q+v) \quad {\rm for} \ w=1 \ , 
\label{lamd1jplusw1}
\eeq
\beq
\lambda_{Z,sq,2,1,-} = v^2 \quad {\rm for} \ w=1 \ .
\label{lamd1jminusw1}
\eeq
The roots of the cubic equation (\ref{eqsqd1cubic}) reduce to $v^2$ and to
$\lambda_{sq,2,1,\pm}$ in Eq. (\ref{lamd1plusminus}), each with multiplicity 
1. 

In the $d=0$ sector, the term (i) reduces to $v(q+v)$ with multiplicity 1 and
(ii) reduces to $v^2$ with multiplicity $s^2-3s+1$.  The
$\lambda_{Z,sq,2,0,\pm}$ reduce to $v(q+v)$ and $v^2$, each with multiplicity
$s-1$. The roots of the cubic equation (\ref{sqlamlad_d0_eq3}) reduce to $v^2$
and $\lambda_{sq,2,1,\pm}$ in Eq. (\ref{lamd1plusminus}), each with
multiplicity $s-1$.  Finally, the roots of the quintic equation reduce to
$v^2$, $\lambda_{sq,2,1,\pm}$ in Eq. (\ref{lamd1plusminus}), and
$\lambda_{sq,2,0,\pm}$ in Eq. (\ref{lamd0plusminus}), each with multiplicity 1.
It is then readily confirmed that in this zero-field case $w=1$, our result
satisfies the identity
\beq
Z(sq, 2 \times m,cyc.,q,s,v,1) = Z(sq,2\times m,cyc.,q,v) \ , 
\label{zhladw1}
\eeq
with $Z(sq,2\times m,cyc.,q,v)$ calculated in \cite{a}, as given in
Eqs. (\ref{zlad})-(\ref{lamd1plusminus}). We have also confirmed that our
result satisfies the various identities given in Sect. \ref{properties} and
that in the special case $s=1$, it reduces to our previous solution of $Z(sq, 2
\times m,cyc.,q,1,v,w)$ given in \cite{zth}.


\section{Partition Function for Cyclic Triangular-Lattice Ladder Graphs}
\label{triladder}

In this section we present our exact results for the partition functions of the
$q$-state Potts model in a generalized magnetic field on cyclic ladder graphs
of the triangular lattice of length $L_x=m$ vertices in the longitudinal
direction, $Z(tri,2 \times m,cyc.,q,s,v,w)$.  As was the case with the
square-lattice ladder strip, the partition function for the triangular-lattice
ladder strip has the form of (\ref{zsumcyc}) with $n_{Zh}(2,2,s)=1$ as a
special case of (\ref{nzhLyLy}), and with the $n_{Zh}(2,d,s)$ given in
Eq. (\ref{nzhLy2d1}) for $d=1$ and in Eq. (\ref{nzhLy2d0}) for $d=0$. Our
result generalizes the zero-field solution given in \cite{ta} and the solution
for the special case $s=1$ given in \cite{zth}

We calculate the $\lambda_{Z,tri,2,d,j}s$ to be as follows:

\begin{itemize}

\item 

For $d=2$, there is a single term, 
\beq
\lambda_{Z,tri,2,2,1} \equiv \lambda_{Z,tri,2,2} = v^2 \ ,
\label{trilamlad_d2}
\eeq
as the $L_y=2$ special case of (\ref{lamladLyLy}); 

\item 

For the $d=1$ sector, i.e., the $\lambda_{Z,tri,2,d,j}$s with coefficient
$\tilde c^{(1)}(q)$, there are $2s+3$ terms. We find the following results: (i)
terms equal to $v^2w$, with multiplicity $2(s-1)$, which thus may be denoted as
\beq
\lambda_{Z,tri,2,1,j}=\lambda_{Z,sq,2,1,j} = 
v^2w \quad {\rm for} \ 1 \le j \le 2(s-1) \ , 
\label{trilamlad_d1_eq1}
\eeq
(ii) next, five terms $\lambda_{Z,tri,2,1,j}$ with $2s-1 \le j \le 2s+3$ 
which are the roots of a quintic equation that is the characteristic polynomial
of the transfer matrix $T_{Z,tri,2,1}$ given in Appendix 
\ref{appendix_tztrid10}.

\item 

For $d=0$, there are $s^2+2s+2$ terms $\lambda_{Z,tri,2,0,j}$. Of these, one
set coincides with the corresponding set for the square-lattice strip, namely
(i): $(vw)^2$, each occurring with multiplicity $s^2-3s+1$. The remaining terms
include (ii) five roots of a quintic equation which is the characteristic
polynomial of the transfer matrix $T_{Z,tri,2,0a}$ (see Appendix
\ref{appendix_tztrid10}), each with multiplicity $s-1$, and (iii) six terms
which are the roots of a degree-6 equation that is the characteristic
polynomial of the transfer matrix $T_{Z,tri,2,0b}$ given in Appendix
\ref{appendix_tztrid10}, each with multiplicity 1. 

\end{itemize}
The reductions in the case $H=0$ (i.e., $w=1$) are similar to those for the
square-lattice strip, so we do not discuss them in detail. 


\section{Some Thermodynamic Properties}
\label{therm}


\subsection{Thermodynamic Functions}
\label{free_energy}

For a lattice strip graph $G_m$ of length $L_x=m$ vertices with $n=n(G_m)$
vertices, we denote the formal limit of infinite length as 
\beq
\lim_{m \to \infty} G_m \equiv \{G\} \ .
\label{gminf}
\eeq
The corresponding dimensionless reduced free energy $f$ in the limit $m \to
\infty$ and hence $n=m L_y \to \infty$ as
\beq
f(\{G\},q,s,v,w) = \lim_{n \to \infty} n^{-1} \, Z(G_m,q,s,v,w) \ , 
\label{f}
\eeq
so that the actual Gibbs free energy is ${\cal G}(T,H)=-k_BT \, f$. For the
$G_m = \Lambda, 2 \times m,cyc.$ cyclic strip graphs considered here, in the $m
\to \infty$ limit, only the $\lambda_{Z,\Lambda,2,d,j}$ of maximum magnitude
contributes.  This comes from the set of terms with $d=0$
\cite{a,ta,ph,phs,s3a}. This is in accord with the property that $f$ should be
independent of the (longitudinal) boundary condition and the fact that only the
$d=0$ sector is present for free longitudinal boundary conditions.  

For the square-lattice ladder strip, among the respective $s^2+2s+2$ terms with
$d=0$ that contribute to Eq. (\ref{zsumcyc}), we find that the dominant one is
the term of largest magnitude among the five roots of the characteristic
polynomial of the transfer matrix $T_{Z,sq,2,0}$ given in Appendix
\ref{appendix_tzsqd0}.  For the triangular-lattice ladder strip, among the
respective $s^2+2s+2$ terms with $d=0$ that contribute to Eq. (\ref{zsumcyc}),
we find that the dominant one is the term of largest magnitude among the six
roots of the characteristic polynomial of the transfer matrix $T_{Z,tri,2,0}$
(see Appendix \ref{appendix_tztrid10}).

Thermodynamic functions are obtained from the reduced free energy (\ref{f}) in
the usual way. For first derivatives, the internal energy per vertex is
$U = -\partial f/\partial \beta = -Jy \, \partial f/\partial y$ and a 
magnetization per vertex
\beq
M = \frac{\partial f}{\partial h} = w\frac{\partial f}{\partial w} \ .
\label{m}
\eeq
It is conventional to define a corresponding magnetic order parameter per 
vertex as (e.g., \cite{wurev})
\beq
{\cal M} = \frac{qM-1}{q-1} \ . 
\label{calm}
\eeq
In the zero-field case, with this definition, in the high-temperature phase,
where the global $S_q$-symmetry is realized manifestly, the spins take on 
random values in the set $I_q$, so $M=1/q$ and hence ${\cal M}=0$.  In
the case of complete ferromagnetic ordering, $M=1$, and hence also 
${\cal M}=1$. 


\subsection{Cases where Frustration and Competing Interactions Occur} 
\label{frustration}

One of the interesting features of our results is that they provide exactly
solved examples that exhibit the occurrence of frustration and competing
interactions as a function of a parameter, $s$, that one can choose.  As
discussed in the introduction, in the antiferromagnetic case with $J < 0$, 
if $H > 0$ (i.e., $w > 1$) and $s < \chi(G)$ or
if $H < 0$ (i.e., $0 \le w < 1$) and $q-s < \chi(G)$, then the imposition of
the respective positive or negative $H$ involves competing interactions and
causes frustration.  Given the symmetry (\ref{zsym}), without loss of
generality, it will suffice to discuss only the case of $H > 0$, and we shall
do so.  For our lattice strip graphs,
\beq
\chi(sq,L_y \times m,cyc.) = \cases{ 2 & if \ $m$ \ is \ even \cr
                                   3 & if \ $m$ \ is \ odd }
\label{chi_sqstrip}
\eeq
\beq
\chi(tri,L_y \times m,cyc.) = \cases{ 3 & if \ $m=0$ \ mod \ 3 \cr
                                      4 & if \ $m=1$ \ or $m=2$ \ mod \ 3 }
\ . 
\label{chi_tristrip}
\eeq
Since the limit $m \to \infty$ can be taken with even $m$ values for the cyclic
square-lattice strip and with values of $m$ that are equal to 0 \ mod \ 3 for
the triangular-lattice strip, and since only the dominant
$\lambda_{Z,\Lambda,L_y,0,j}$ is relevant in this limit, it follows that the
competition and frustration is exhibited in this dominant
$\lambda_{Z,\Lambda,L_y,0,j}$ for $H > 0$ and $0 < s < 2$, i.e., $s=1$ for
cyclic square-lattice strips and $0 < s < 3$, i.e., $s=1$ or $s=2$, for cyclic
triangular-lattice strips.  In each case, in accordance with our discussion in
the introduction, it is assumed that in these respective cases, the chromatic
number (2 or 3) is no greater than $q$, so that the trivial situation where
$s=q$ does not occur. Similarly, there is frustration if $H < 0$ and $0 < q-s <
2$ for square-lattice strips and $0 < q-s < 3$ for triangular-lattice strips.

Our discussion above is general. An explicit example of this phenomenon is
instructive.  The simplest illustration of the frustration is provided by $Z$
for the circuit graph, $Z(C_m,q,s,v,w)$.  Here, the dominant term is
$\lambda_{Z,C,0,+}$ (given in Eqs. (\ref{lamcnd0plusminus})-(\ref{rc})), so
\beq
f(\{C\},q,v,s,w) = \ln[\lambda_{Z,C,0,+}] \ ,
\label{fcn}
\eeq
where, as in Eq. (\ref{gminf}), ${\cal C}$ denotes the formal limit of the
circuit graph $C_m$ as $m \to \infty$.  For our discussion, it will be useful
to have the explicit form of $M$ resulting from Eq. (\ref{fcn}). We calculate
\beq
M = \frac{w}{\sqrt{R}} \, \Bigg [s+v - \frac{2v(q+v)}{A + \sqrt{R}} \ \Bigg ]
\ , 
\label{mcn}
\eeq
where $A$ and $R$ were given in Eqs. (\ref{ac}) and (\ref{rc}).  The magnetic
order parameter is then obtained by substituting this expression for $M$ in
Eq. (\ref{calm}). 

In the limit $m \to \infty$, taken on even $m$, $\chi(\{ C \})=2$.  
If $J < 0$ with $H > 0$, then as $T \to 0$, the spin-spin interaction
requires that the spins on adjacent vertices should be different, while the
magnetic field term biases them to lie in the interval $I_s$. These
requirements conflict with each other if $s=1$, since this is less than 
$\chi(\{ C \})$. In this $s=1$ case, as $T \to 0$, 
\beq
\lambda_{Z,C,0,+} \to \frac{1}{2}\Big [ q-2 + e^{h-|K|} + 
\sqrt{(q-2+e^{h-|K|})^2 + 4(q-1)e^h } \ \ \Big ]
\label{lamcnd0plus_s1}
\eeq
(where we have used the fact that $v \to -1$ in this limit).  Now, if $H/|J| <
1$, i.e., $h/|K| < 1$, then $e^{h-|K|} \to 0$ as $T \to 0$, while $e^h$ grows
without bound, so $\lambda_{Z,C,0,+} \to \sqrt{(q-1)e^h}$ asymptotically, up to
exponentially smaller correction terms.  If $1 < h/|K| < 2$, then, although the
$e^{h-|K|}$ term outside the square root and the $e^{2(h-|K|)}$ term inside the
square root grow, they grow less rapidly than the $4(q-1)e^h$ term inside the
square root, so as $T \to 0$, the dominant term in $\lambda_{Z,C,0,+}$
asymptotically is again $\sqrt{(q-1)e^h}$.  If $h/|K| > 2$, then the
$e^{2(h-|K|)}$ term inside the square root dominates over the $4(q-1)e^h$ term
so, including also the $e^{h-|K|}$ term outside the square root, one finds
that, as $T \to 0$, $\lambda_{Z,C,0,+}$ asymptotically approaches $e^{h-|K|}$,
with exponentially smaller correction terms. These are the generic intervals in
$h/|K|$.  One can also trace this demarcation point $h/|K|=2$ separating the
two types of behavior in the space of complexified parameters, as, e.g., in
another exactly solved spin model exhibiting frustration and competing
interactions \cite{1dnnn}.  For completeness, one should also note the marginal
case where $H=2|J|$, i.e., $h/|K|=2$, which forms a set of measure zero in the
space of the parameters $J$ and $H$; in this case, if one expresses
$\lambda_{Z,C,0,+}$ in terms of $|K|$ alone, then as $T \to 0$,
$\lambda_{Z,C,0,+} \to (1/2)e^{|K|}[1+\sqrt{4q-3} \ ]$.

We now apply this analysis to $M$, as calculated in Eq. (\ref{mcn}) and the
resultant ${\cal M}$ from Eq. (\ref{calm}). In the case that we are
considering, with $J < 0$ and $s=1$, as $T \to 0$, if $H > 2|J|$, the
ferromagnetic tendency due to the external field dominates over the proper
$q$-coloring tendency due to the spin-spin interaction, so
\beq
J < 1, \ H > 2|J|, \ s=1 \ \Rightarrow \ M={\cal M}=1 \ . 
\label{lamcnd0plus_s1_hwins}
\eeq
If, on the other hand, $H < 2|J|$, then
\beq
J < 1, \ H < 2|J|, \ s=1 \ \Rightarrow \ M=\frac{1}{2}, \quad
{\cal M}=\frac{q-2}{2(q-1)} \ . 
\label{mafm}
\eeq
Thus, for $J < 0$ and $s=1$, in the limit $T \to 0$, as $H$ decreases through
the value $2|J|$, there is a discontinuous decrease in ${\cal M}$ from 1 to the
value in Eq. (\ref{mafm}).  For $q=2$, this latter value is 0; for $q=3$, it is
1/4, and as $q \to \infty$, it approaches 1/2 from below. In this situation
with frustration and competing interactions, there is thus a nonanalytic change
in the magnetization at $T=0$ as a consequence of a change in the ratio
$H/|J|$. Our example has the appeal that the physics is manifestly evident
because of the explicit exact solution. 


\section{Conclusions}
\label{conclusions}

In this paper we have presented an exact calculations of the partition
functions $Z(\Lambda,2 \times m;q,s,v,w)$ of the $q$-state Potts model in a
generalized magnetic field, on cyclic ladder strip graphs of the square and
triangular lattices of arbitrary length. Several interesting analytic
properties of these results have been elucidated. In particular, we have shown
that, for the case of antiferromagnet spin-spin coupling, these provide exactly
solved models that exhibit an onset of frustration and competing interactions,
with a resultant nonanalytic change in physical quantities at zero temperature,
in the context of a novel type of tensor-product $S_s \otimes S_{q-s}$ global
symmetry.


\begin{acknowledgments}

This research was partly supported by the Taiwan Ministry of Science and
Technology grant MOST 103-2918-I-006-016 (S.-C.C.) and by
the U.S. National Science Foundation grant No. NSF-PHY-13-16617 (R.S.).

\end{acknowledgments}


\begin{appendix}


\section{$Z(sq,2 \times m,cyc.,q,v)$} 
\label{appendix_zlad}

We review here the result for the partition function $Z(sq,2 \times
m,cyc.,q,v)$ of the cyclic square-lattice ladder graph of length $L_x=m$
vertices presented in \cite{a}. We include this in connection with our
discussion in the text showing how our new result for $Z(sq,2 \times
m,cyc.,q,s,v,w)$ reduces to $Z(sq,2 \times m,cyc.,q,v)$ in the zero-field case
$w=1$. This partition function has the form of 
\beq
Z(sq,L_y \times m,cyc.,q,v) = \sum_{d=0}^{L_y} c^{(d)}
\sum_{j=1}^{n_Z(L_y,d)} (\lambda_{sq,L_y,d,j})^m \ , 
\label{zladgen}
\eeq
with $L_y=2$, where $n_Z(2,0)=2$, $n_Z(2,1)=3$, $n_Z(2,2)=1$, and 
the coefficients $c^{(d)}$ are given in Eq. (\ref{cdtilde}), so
\beq
c^{(0)}=1, \quad c^{(1)}=q-1, \quad c^{(2)}=q^2-3q+1 \ . 
\label{cdlad'}
\eeq
As is evident in Eq. (\ref{zladgen}), to distinguish the $\lambda_{sq,2,d,j}$s
in the zero-field partition function $Z(sq,2 \times m,q,v)$ from the
$\lambda_{Z,sq,2,d,j}$ in the field-dependent partition function $Z(sq,2\times
m,q,s,v,w)$, we suppress the subscript $Z$ in the former. Explicitly, 
\beqs
& & Z(sq,2 \times m,cyc.,q,v)=(\lambda_{sq,2,0,+})^m+(\lambda_{sq,2,0,-})^m
 \cr\cr
& + & c^{(1)}\Big [ (\lambda_{sq,2,1,1})^m + 
(\lambda_{sq,2,1,+})^m + (\lambda_{sq,2,1,-})^m \Big ] + 
c^{(2)} (\lambda_{sq,2,2})^m \ , 
\label{zlad}
\eeqs
where (in order of decreasing $d$) $\lambda_{sq,2,2}=v^2$, 
\beq
\lambda_{sq,2,1,1}=v(q+v) \ , 
\label{lamd1j1}
\eeq
\beqs
\lambda_{sq,2,1,(2,3)} & \equiv & \lambda_{sq,2,1,\pm} \cr\cr
& = & \frac{v}{2}\Big [ q+v(v+4) \pm
(v^4+4v^3+12v^2-2qv^2+4qv+q^2)^{1/2} \Big ] \ , 
\label{lamd1plusminus}
\eeqs
and
\beq
\lambda_{sq,2,0,(1,2)} \equiv \lambda_{sq,2,0,\pm} 
= \frac{1}{2}(A_{sqd0} \pm \sqrt{R_{sqd0}} \ ) \ , 
\label{lamd0plusminus}
\eeq
where
\beq
A_{sqd0}=v^3+4v^2+3qv+q^2
\label{td0}
\eeq
and
\beq
R_{sqd0}=v^6+4v^5-2qv^4-2q^2v^3+12v^4+16qv^3+13q^2v^2+6q^3v+q^4 \ .
\label{rd0}
\eeq
The reader is referred to \cite{ta} for our corresponding solution for the
partition function $Z(tri,2 \times m,cyc.,q,v)$ of the cyclic 
triangular-lattice ladder graph of arbitrary length. 


\section{$T_{Z,sq,2,0}$}
\label{appendix_tzsqd0}

Five of the $s^2+2s+2$ \ $\lambda_{Z,sq,2,0,j}$ terms, each with multiplicity
1, are determined as the roots of a quintic equation which is the
characteristic polynomial of the transfer matrix $T_{Z,sq,L_y,d}$ with $L_y=2$,
$d=0$, and the following entries:
\beq
(T_{Z,sq,2,0})_{1,1} =\tilde q^2 + 3v(\tilde q + v)
\label{tsq11}
\eeq
\beq
(T_{Z,sq,2,0})_{1,2} = 2s(\tilde q + v)w
\label{tsq12}
\eeq
\beq
(T_{Z,sq,2,0})_{1,3} = s(v+1)w^2
\label{tsq13}
\eeq
\beq
(T_{Z,sq,2,0})_{1,4} = s(s-1)w^2
\label{tsq14}
\eeq
\beq
(T_{Z,sq,2,0})_{1,5} = (\tilde q + 2v)(v+1) 
\label{tsq15}
\eeq
%
%
\beq
(T_{Z,sq,2,0})_{2,1} = \tilde q^{\ 2} + v(2\tilde q + v) 
\label{tsq21}
\eeq
\beq
(T_{Z,sq,2,0})_{2,2} = w[2s\tilde q + v(q+v)]
\label{tsq22}
\eeq
\beq
(T_{Z,sq,2,0})_{2,3} = (v+1)(v+s)w^2
\label{tsq23}
\eeq
\beq
(T_{Z,sq,2,0})_{2,4} = (v+s)(s-1)w^2
\label{tsq24}
\eeq
\beq
(T_{Z,sq,2,0})_{2,5} = (\tilde q + v)(v+1) 
\label{tsq25}
\eeq
%
%
\beq
(T_{Z,sq,2,0})_{3,1} =(T_{Z,sq,2,0})_{4,1} = \tilde q ( \tilde q + v) 
\label{tsq31}
\eeq
\beq
(T_{Z,sq,2,0})_{3,2} =(T_{Z,sq,2,0})_{4,2} = 2\tilde q (v+s)w 
\label{tsq32}
\eeq
\beq
(T_{Z,sq,2,0})_{3,3} = (v+1)(v^2+2v+s)w^2
\label{tsq33}
\eeq
\beq
(T_{Z,sq,2,0})_{3,4} = (2v+s)(s-1)w^2
\label{tsq34}
\eeq
\beq
(T_{Z,sq,2,0})_{3,5} = (T_{Z,sq,2,0})_{4,5} = \tilde q (v+1) 
\label{tsq35}
\eeq
%
%
\beq
(T_{Z,sq,2,0})_{4,3} = (v+1)(2v+s)w^2 
\label{tsq43}
\eeq
\beq
(T_{Z,sq,2,0})_{4,4} = [(s+v)^2-s-2v]w^2
\label{tsq44}
\eeq
%
%
\beq
(T_{Z,sq,2,0})_{5,1} = v^3
\label{tsq51}
\eeq
\beq
(T_{Z,sq,2,0})_{5,j} = 0 \quad {\rm for} \ 2 \le j \le 4
\label{tsq5j}
\eeq
\beq
(T_{Z,sq,2,0})_{5,5} = v^2(v+1) 
\label{tsq55}
\eeq
%


\section{Matrices $T_{Z,tri,2,d}$ for $d=1$, $d=0$}
\label{appendix_tztrid10}

\subsection{$d=1$}

Five of the $s^2+2s+2$ \ $\lambda_{Z,tri,2,1,j}$ terms, each with multiplicity
1, are determined as the roots of a quintic equation which is the
characteristic polynomial of the matrix $T_{Z,tri,L_y,d}$ with $L_y=2$, $d=1$,
and the following entries:
\beq
(T_{Z,tri,2,1})_{1,1} = v \tilde q + 2v^2
\label{ttx11}
\eeq
\beq
(T_{Z,tri,2,1})_{1,2} = swv
\label{ttx12}
\eeq
\beq
(T_{Z,tri,2,1})_{1,3} = (T_{Z,tri,2,1})_{2,3} = v^2
\label{ttx13}
\eeq
\beq
(T_{Z,tri,2,1})_{1,4} = (T_{Z,tri,2,1})_{2,4} = 0
\label{ttx14}
\eeq
\beq
(T_{Z,tri,2,1})_{1,5} = (T_{Z,tri,2,1})_{2,5} = v(1+v)
\label{ttx15}
\eeq
%
%
\beq
(T_{Z,tri,2,1})_{2,1} = v \tilde q + v^2
\label{ttx21}
\eeq
\beq
(T_{Z,tri,2,1})_{2,2} = wv(v+s)
\label{ttx22}
\eeq
%
%
\beq
(T_{Z,tri,2,1})_{3,1} = (T_{Z,tri,2,1})_{4,1} = v \tilde q + 3v^2 + v^3
\label{ttx31}
\eeq
\beq
(T_{Z,tri,2,1})_{3,2} = (T_{Z,tri,2,1})_{4,2} = swv 
\label{ttx32}
\eeq
\beq
(T_{Z,tri,2,1})_{3,3} = v \tilde q + 4v^2 + v^3
\label{ttx33}
\eeq
\beq
(T_{Z,tri,2,1})_{3,4} = swv
\label{ttx34}
\eeq
\beq
(T_{Z,tri,2,1})_{3,5} = (T_{Z,tri,2,1})_{4,5} = v(1+v)(2+v)
\label{ttx35}
\eeq
%
%
\beq
(T_{Z,tri,2,1})_{4,3} = v \tilde q + 3v^2 + v^3
\label{ttx43}
\eeq
\beq
(T_{Z,tri,2,1})_{4,4} = wv(v+s)
\label{ttx44}
\eeq
%
%
\beq
(T_{Z,tri,2,1})_{5,1} = v^2 \tilde q + 3v^3 + v^4
\label{ttx51}
\eeq
\beq
(T_{Z,tri,2,1})_{5,2} = swv^2
\label{ttx52}
\eeq
\beq
(T_{Z,tri,2,1})_{5,3} = 3v^3 + v^4
\label{ttx53}
\eeq
\beq
(T_{Z,tri,2,1})_{5,4} = 0
\label{ttx54}
\eeq
\beq
(T_{Z,tri,2,1})_{5,5} = v^2(1+v)(2+v)
\label{ttx55}
\eeq


\subsection{$d=0$}

Five of the $\lambda_{Z,tri,L_y,d,j}$ with $L_y=2$ and $d=0$ are the roots,
each with multiplicity $s-1$, of a quintic equation which is the characteristic
polynomial of the transfer matrix $T_{Z,tri,2,0a}$. This matrix may be obtained
from $T_{Z,tri,2,1}$ by the replacements $s \to q-s$ and $w \to w^{-1}$ and
then multiplication by $w^2$. 

Six of the $\lambda_{Z,tri,L_y,d,j}$ with $L_y=2$ and $d=0$ are the roots, each
with multiplicity 1, of a degree-6 equation which is the characteristic
polynomial of the matrix $T_{Z,tri,2,0b}$ with entries
\beq
(T_{Z,tri,2,0b})_{1,1} = \tilde q^2 + 4v\tilde q + 5v^2 + v^3
\label{ttu11}
\eeq
\beq
(T_{Z,tri,2,0b})_{1,2} = sw(\tilde q + v)
\label{ttu12}
\eeq
\beq
(T_{Z,tri,2,0b})_{1,3} = sw(\tilde q + 2v)
\label{ttu13}
\eeq
\beq
(T_{Z,tri,2,0b})_{1,4} = sw^2(1+v)
\label{ttu14}
\eeq
\beq
(T_{Z,tri,2,0b})_{1,5} = s(s-1)w^2
\label{ttu15}
\eeq
\beq
(T_{Z,tri,2,0b})_{1,6} = (1+v)(\tilde q + 3v + v^2)
\label{ttu16}
\eeq
%
%
\beq
(T_{Z,tri,2,0b})_{2,1} = \tilde q^2 + 3v\tilde q + 3v^2 + v^3
\label{ttu21}
\eeq
\beq
(T_{Z,tri,2,0b})_{2,2} = w(v+s)(\tilde q + v)
\label{ttu22}
\eeq
\beq
(T_{Z,tri,2,0b})_{2,3} = sw(\tilde q + v)
\label{ttu23}
\eeq
\beq
(T_{Z,tri,2,0b})_{2,4} = w^2(1+v)(v+s)
\label{ttu24}
\eeq
\beq
(T_{Z,tri,2,0b})_{2,5} = w^2(s-1)(v+s)
\label{ttu25}
\eeq
\beq
(T_{Z,tri,2,0b})_{2,6} = (1+v)(\tilde q + 2v + v^2)
\label{ttu26}
\eeq
%
%
\beq
(T_{Z,tri,2,0b})_{3,1} = \tilde q^2 + 2v\tilde q + v^2
\label{ttu31}
\eeq
\beq
(T_{Z,tri,2,0b})_{3,2} = w\tilde q(v + s)
\label{ttu32}
\eeq
\beq
(T_{Z,tri,2,0b})_{3,3} = w(v+s)(\tilde q + v)
\label{ttu33}
\eeq
\beq
(T_{Z,tri,2,0b})_{3,4} = w^2(1+v)(s+2v+v^2)
\label{ttu34}
\eeq
\beq
(T_{Z,tri,2,0b})_{3,5} = w^2(2vs-2v-s+s^2)
\label{ttu35}
\eeq
\beq
(T_{Z,tri,2,0b})_{3,6} = (1+v)(\tilde q + v)
\label{ttu36}
\eeq
%
%
\beq
(T_{Z,tri,2,0b})_{4,1} = (T_{Z,tri,2,0b})_{5,1} = \tilde q (\tilde q + v)
\label{ttu41}
\eeq
\beq
(T_{Z,tri,2,0b})_{4,2} = w\tilde q (s + 2v + v^2)
\label{ttu42}
\eeq
\beq
(T_{Z,tri,2,0b})_{4,3} = (T_{Z,tri,2,0b})_{5,3} = w\tilde q (s+v)
\label{ttu43}
\eeq
\beq
(T_{Z,tri,2,0b})_{4,4} = w^2(1+v)(s + 3v + 3v^2 + v^3)
\label{ttu44}
\eeq
\beq
(T_{Z,tri,2,0b})_{4,5} = w^2(s-1)(s+3v+v^2)
\label{ttu45}
\eeq
\beq
(T_{Z,tri,2,0b})_{4,6} = (T_{Z,tri,2,0b})_{5,6} = \tilde q (1+v)
\label{ttu46}
\eeq
%
%
\beq
(T_{Z,tri,2,0b})_{5,2} = w\tilde q (s+2v)
\label{ttu52}
\eeq
\beq
(T_{Z,tri,2,0b})_{5,4} = w^2(1+v)(s+3v+v^2)
\label{ttu54}
\eeq
\beq
(T_{Z,tri,2,0b})_{5,5} = w^2(s^2-s+3vs-3v+v^2)
\label{ttu55}
\eeq
%
%
\beq
(T_{Z,tri,2,0b})_{6,1} = v^2 (\tilde q + 3v + v^2) 
\label{ttu61}
\eeq
\beq
(T_{Z,tri,2,0b})_{6,2} = (T_{Z,tri,2,0b})_{6,4} = (T_{Z,tri,2,0b})_{6,5} = 0
\label{ttu62}
\eeq
\beq
(T_{Z,tri,2,0b})_{6,3} = swv^2
\label{ttu63}
\eeq
\beq
(T_{Z,tri,2,0b})_{6,6} = v^2(1+v)(2+v)
\label{ttu66}
\eeq
%


\end{appendix}


\end{document}